\documentclass[12pt]{iopart}
\usepackage{graphicx}

\begin{document}

\title{Temperature oscillations of magnetization observed in nanofluid
ferromagnetic graphite}
\author{S Sergeenkov$^{1,3}$,  N S Souza$^{1}$, C Speglich$^{1}$, V A G
Rivera$^{1}$, C A Cardoso$^{1}$, H Pardo$^{2}$, A W
Mombr\'{u}$^{2}$ and F M Ara\'{u}jo-Moreira$^{1,4}$}

\address{$^{1}$Materials and Devices Group, Department of
Physics and Physical Engineering, Universidade Federal
de S\~ao Carlos, S\~ao Carlos, SP, 13565-905 Brazil\\
$^{2}$Crystallography, Solid State and Materials Laboratory
(Cryssmat-Lab), DEQUIFIM, Facultad de Qu\'{i}mica, Universidad de
la Rep\'{u}blica, P.O. Box 1157, CP 11800, Montevideo, Uruguay\\
$^{3}$Corresponding author, E-mail: sergei@df.ufscar.br\\
$^{4}$Research Leader, E-mail: faraujo@df.ufscar.br}

\begin{abstract}
We report on unusual magnetic properties observed in the nanofluid
room-temperature ferromagnetic graphite (with an average particle
size of $l\simeq 10nm$). More precisely, the measured
magnetization exhibits a low-temperature anomaly (attributed to
manifestation of finite size effects below the quantum temperature
$T_l\propto \hbar^2/l^2$) as well as pronounced temperature
oscillations above $T=50K$ (attributed to manifestation of the
hard-sphere type pair correlations between ferromagnetic particles
in the nanofluid).
\end{abstract}

\pacs{75.50Dd, 75.50Mm, 75.50.Tt, 81.05.Uw, 82.70.Kj}

\maketitle

Recently, quite a substantial progress has been made in developing
suspended colloids of nano sized magnetic particles, including
carbon, graphite and graphene based nanofluids and biocompatible
ferrofluids (see, e.g.,[1-9] and further references therein). In
particular, Parkansky {\it et al} [6] successfully separated
magnetic carbon particles (including chains of nanospheres with
diameters from $30$ to $50nm$, and nanorods with lengths from $50$
to $250nm$ and diameters from $20$ to $30nm$) in the obtained
solutions by means of the bio-ferrography technique. At the same
time, Widenkvist {\it et al} [9] suggested a new method to produce
suspensions of graphene sheets (graphite flakes) by combining
solution-based bromine intercalation and mild sonochemical
exfoliation.

In this paper, we report on the magnetic properties of the
nanofluid magnetic graphite (NFMG) obtained from the previously
synthesized bulk organic magnetic graphite (MG) by stabilizing the
aqueous ferrofluid suspension with an addition of active cationic
surfactant. Two interesting phenomena have been observed in the
temperature behavior of the magnetization: a low-temperature
anomaly (attributed to the manifestation of quantum size effects
due to an average particle size of the order of $l\simeq 10nm$),
and pronounced temperature oscillations above $T=50K$ (attributed
to manifestation of the hard-sphere type pair correlations between
ferromagnetic particles in the nanofluid).

\begin{figure}
\centerline{\includegraphics[width=14.0cm,angle=0]{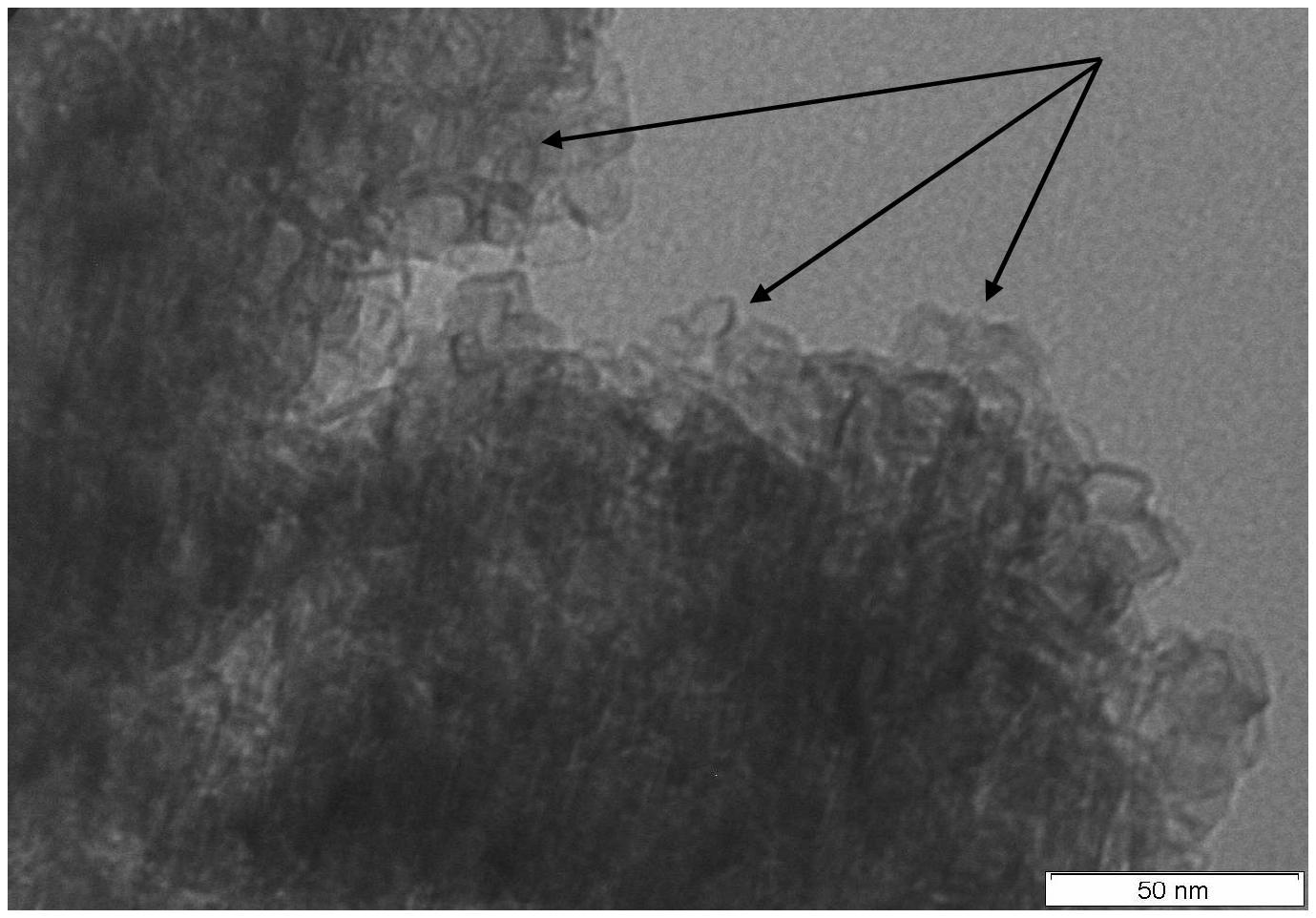}}
\vspace{0.25cm} \caption{TEM image of NFMG sample. }
\label{fig:fig1}
\end{figure}

Recall [10-13] that the chemically modified magnetic graphite (MG)
was produced by a vapor phase redox controlled reaction in closed
nitrogen atmosphere with addition of copper oxide using synthetic
graphite powder. The obtained in such a way modified graphite has
a strong magnetic response even at room temperature (which
manifests itself through a visible attraction by a commercial
magnet). After obtaining the MG, we have prepared the nanofluid
suspension (NFMG) by dissolving graphite in acetone, adding a
Cetyltrimethylammonium bromide (CTAB) cationic surfactant, and
bringing it to an ultra sonic edge. The structural
characterization of NFMG was performed by transmission electron
microscopy (TEM) using Philips CM-120 microscope. The analysis of
TEM images (shown in Fig.\ref{fig:fig1}) reveals clusters (ranging
from $100$ to $300nm$) with an average size of the ferromagnetic
particle in the nanofluid of the order of $10nm$ (more details
regarding the structure and chemical route for synthesis of the
discussed here nanofluid magnetic graphite will be presented
elsewhere [14]).

\begin{figure}
\centerline{\includegraphics[width=9.0cm]{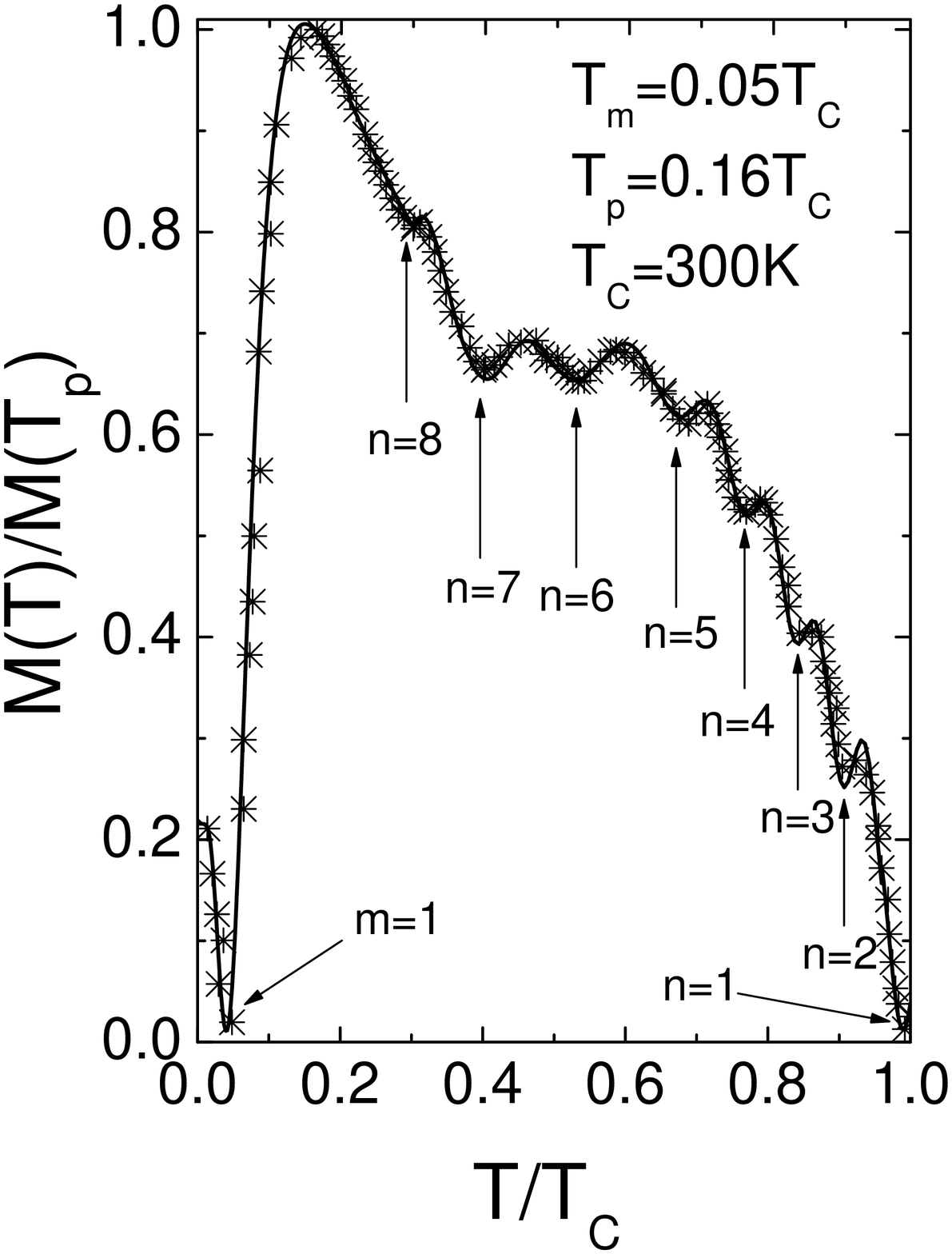}}\vspace{0.25cm}
\caption{The temperature dependence of the normalized
magnetization of NFMG (after subtracting the paramagnetic
contribution). The solid line is the best fit according to
Eqs.(1)-(3).} \label{fig:fig2}
\end{figure}

\begin{figure}
\centerline{\includegraphics[width=8.0cm,angle=0]{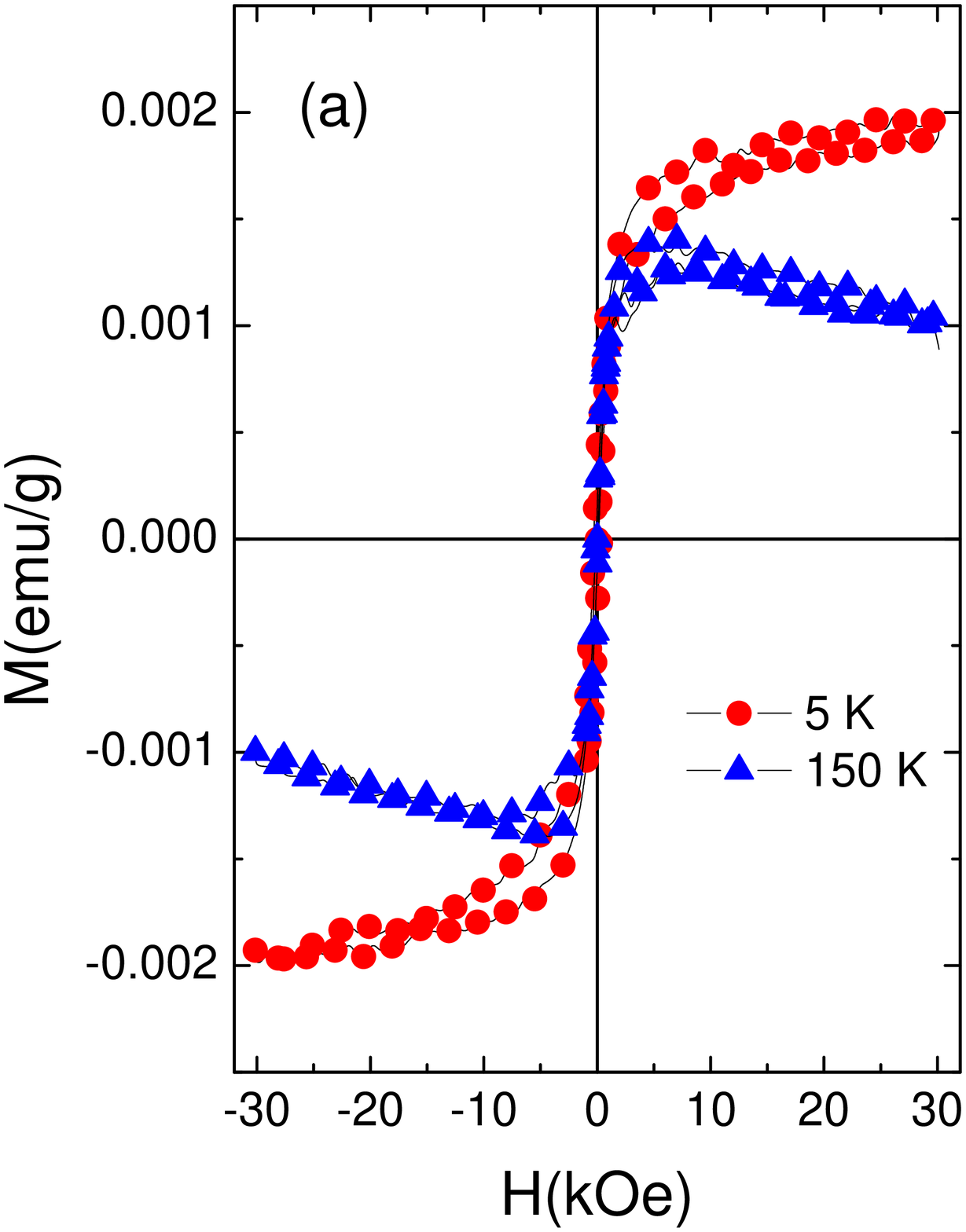}}\vspace{0.5cm}
\centerline{\includegraphics[width=8.0cm,angle=0]{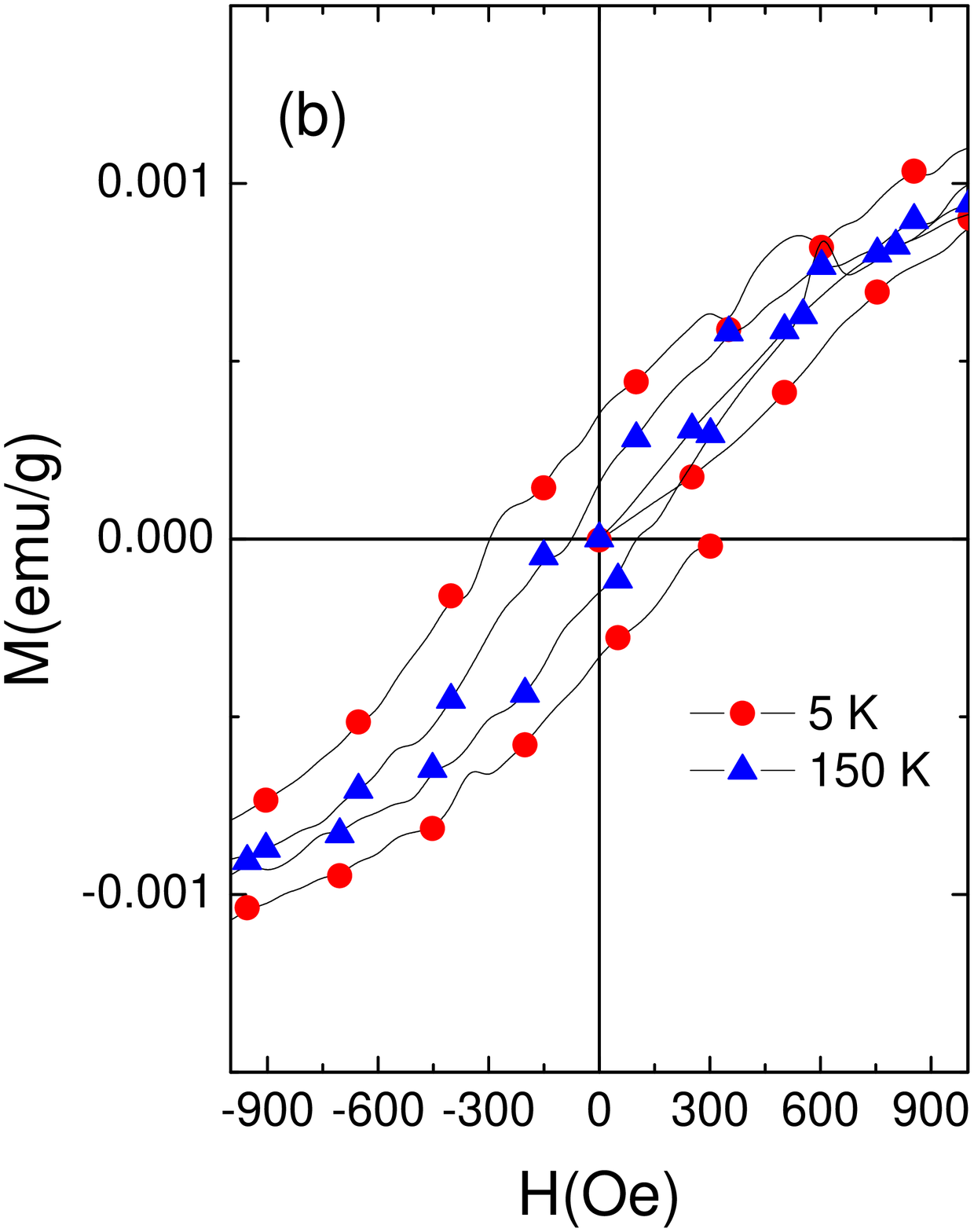}}
\vspace{0.05cm} \caption{The hysteresis curves (taken at two
temperatures, $5K$ and $150K$) for high (a) and low (b) applied
magnetic field regions, showing a ferromagnetic like behavior of
the NFMG sample.} \label{fig:fig3}
\end{figure}

To test the magnetic properties of NFMG samples, we performed the
standard zero field cooled (ZFC) and field cooled (FC)
measurements using a MPMS-5T SQUID magnetometer from Quantum
Design. Fig.\ref{fig:fig2} presents the temperature dependence of
the normalized magnetization $M(T)/M(T_p)$ (taken under the
applied magnetic field of $1 kOe$) after subtraction of
paramagnetic contributions ($T_p=0.16T_C=48K$ is the temperature
where $M(T)$ has a maximum with the absolute value of
$M(T_p)=0.1emu/g$). Notice that there are two distinctive regions,
below and above the peak temperature $T_p$. Namely, below $T_p$
there is a well-defined low-temperature minimum (around
$T_{m}=0.05T_C=15K$), while for $T>T_p$ we have pronounced
temperature oscillations. To verify that the observed peak
originates from true quantum effects in ferromagnetic sample
(rather than from superparamagnetic behavior due to the thermal
energy domination over anisotropy energy), we also measured the
hysteretic $M-H$ curves for two characteristic temperatures,
$T=5K<T_m$ (in the region of fully fledged quantum effects) and
$T=150K=0.5T_C$ (in the middle of the oscillations pattern).
According to Fig.\ref{fig:fig3}, the low-temperature hysteresis is
quite strong (with coercive magnetic field $H_C=338Oe$) and it
does not disappear with increasing the temperature ($H_C=200 Oe$
for $T=150K$). Thus, we can safely assume that the observed
temperature features in our sample do originate from a true
ferromagnetic behavior.

Turning to the analysis of the obtained results, let us begin with
the low-temperature region ($T<T_p$) and discuss the origin of the
observed minimum of magnetization near $T_m=0.05T_C$. Recall that
the finite temperature quantum effects manifest themselves for the
size of the particle $l<\Lambda (T)$ (where $\Lambda
(T)=\sqrt{2\pi \hbar^2/m^{*}k_BT}$ is the thermal de Broglie
wavelength) or, alternatively, for temperatures $T<T_l$ (where
$T_l=2\pi \hbar^2/m^{*}k_Bl^2$ is the quantum temperature). Using
$l\simeq 10nm$ for an average size of the single particle in our
samples (and assuming free electron mass for $m^{*}$), we get
$T_l=0.15T_C= 45K$ for the onset temperature below which the
manifestation of quantum size effects is expected (notice that
$T_l$ is very close to the peak temperature $T_p=0.16T_C$). To fit
the low-temperature experimental data, we assume the following
normalized (to the peak temperature $T_p$) periodic dependence of
the finite-size magnetization:

\begin{equation}
\frac{M_l(T)}{M_l(T_p)}=\left[\frac{l}{\Lambda(T)}\right]\sin\left\{\left[
\frac{M_{\infty}(T)}{M_{\infty}(T_p)}\right]\left[\frac{\Lambda(T)}{l}\right]\right\}
\end{equation}
where $M_{\infty}(T)$ is the bulk magnetization of a single
magnetic particle.

It can be easily verified that Eq.(1) reduces to $M_{\infty}(T)$
when the quantum effects become negligible. More precisely,
$M_{\infty}(T)/M_{\infty}(T_p)=\lim_{l\gg
\Lambda(T)}[M_l(T)/M_l(T_p)]$.

We were able to successfully fit the low-temperature data using
the following explicit expression for the single particle bulk
magnetization:
\begin{equation}
M_{\infty}(T)=M_{s}\tanh{\sqrt{\left(\frac{T_C}{T}\right)^2-1}}
\end{equation}
which presents analytical (approximate) solution of the
Curie-Weiss mean-field equation for spontaneous magnetization
valid for all temperatures [15,16]. The solid line in
Fig.\ref{fig:fig2} presents the best fits for low-temperature
region ($T\le T_p$) according to Eqs.(1) and (2) with
$M_{s}=0.95M(T_p)$, $M(T_p)=0.1emu/g$, $T_C=300K$ and $l=10nm$.

\begin{figure}
\centerline{\includegraphics[width=8.50cm]{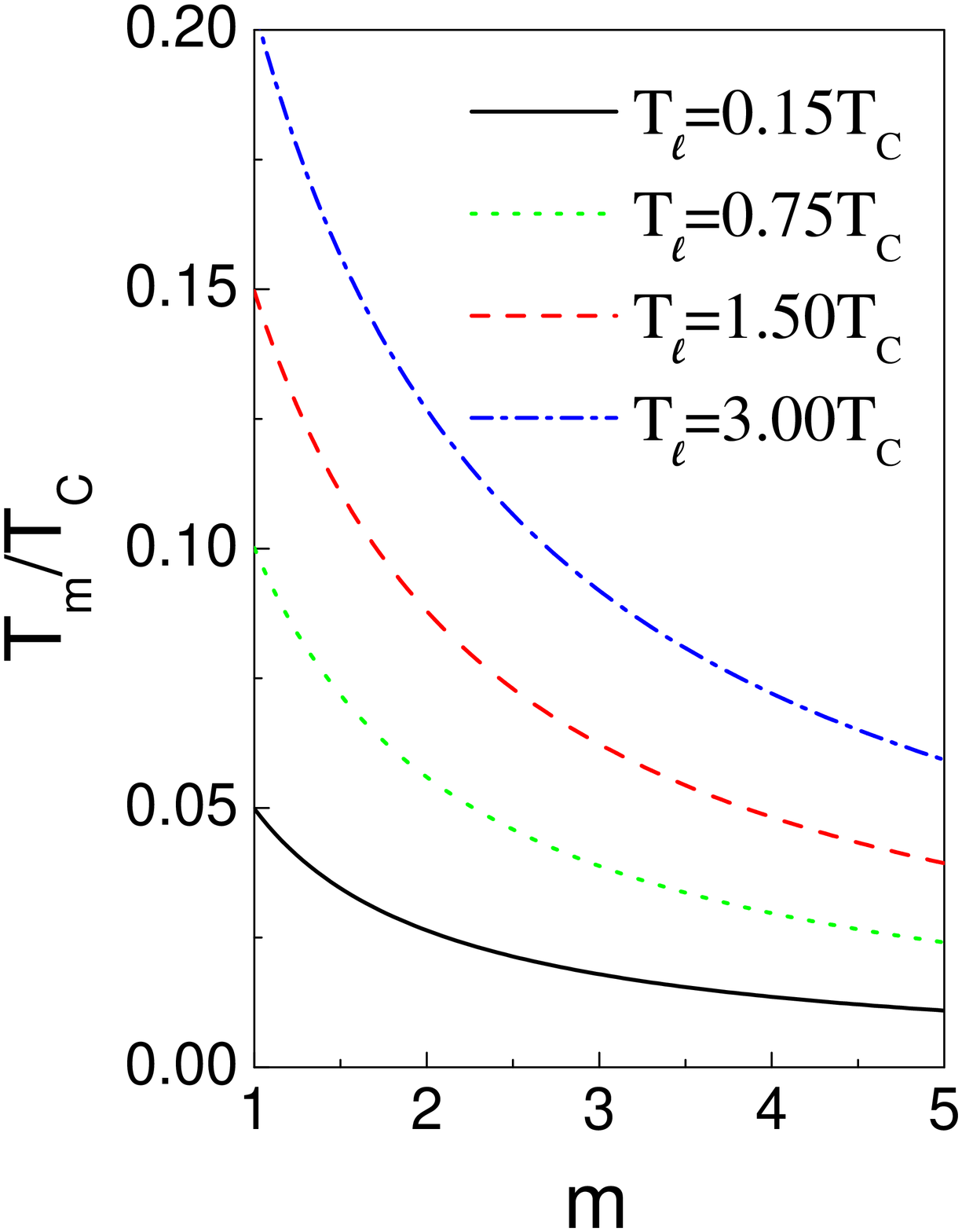}}\vspace{0.25cm}
\caption{The predicted dependence of the reduced temperature
$T_m/T_C$ on the oscillations minima $m$ for different values of
the particle size $l$ related quantum temperature
$T_l$.}\label{fig:fig4}
\end{figure}

Notice also that, for a given temperature, the above periodic
function $M_l(T)$ has minima $m$ at $T=T_m$ where $T_m$ is the
solution of the implicit equation, $M_{\infty}(T_m)\Lambda
(T_m)=\pi mM_{\infty}(T_p)l$ with $m=1,2,...$ being the number of
the oscillation minima. Using the Curie-Weiss expression for bulk
magnetization $M_{\infty}(T)$ and previously defined thermal de
Broglie wavelength $\Lambda(T)$, in Fig.\ref{fig:fig4} we depict
the solution of the above equation as the dependence of the
reduced temperature $T_m/T_C$ on $m$ for different values of the
particle size $l$ (in terms of the quantum temperature $T_l\propto
\hbar^2/l^2$). According to this picture, the smaller the particle
size (hence, the larger the quantization temperature $T_l$), the
more finite size related oscillations (minima) should be observed
in the temperature dependence of the magnetization $M_l(T)$. For
example, in our particular case (with $l=10nm$ and $T_l=0.15T_C$)
only first minimum ($m=1$) is expected to be visible at non-zero
temperature $T_m=0.05T_C=15K$, in agreement with the observations
(see Fig.\ref{fig:fig2}).

Turning to the discussion of the high-temperature region (above
$T_p$), it is quite natural to assume that the observed
oscillations can be attributed to the local variation of the
magnetization ${\tilde M}_f(r)=M_0g(r)$ defined by the periodic
radial distribution function $r^2g(r)=\sin kr$ in the hard-sphere
fluid model [17,18] with $k(T)=\pi M_{\infty}(T)/M_{\infty}(T_p)l$
where $M_{\infty}(T)$ is the above-introduced bulk magnetization
of the single particle. Within this scenario, the temperature
dependence of the fluid contribution to magnetization reads:
\begin{equation}
M_f(T)=\frac{1}{L}\int_0^{L}r^2dr {\tilde
M}_f(r)=M_0\left[\frac{1-\cos k(T)L}{k(T)L}\right]
\end{equation}
where $L=N_fl$  and $M_0=[\pi N_f/(1-\cos\pi N_f)]M_f(T_p)$.

\begin{figure}
\centerline{\includegraphics[width=8.50cm]{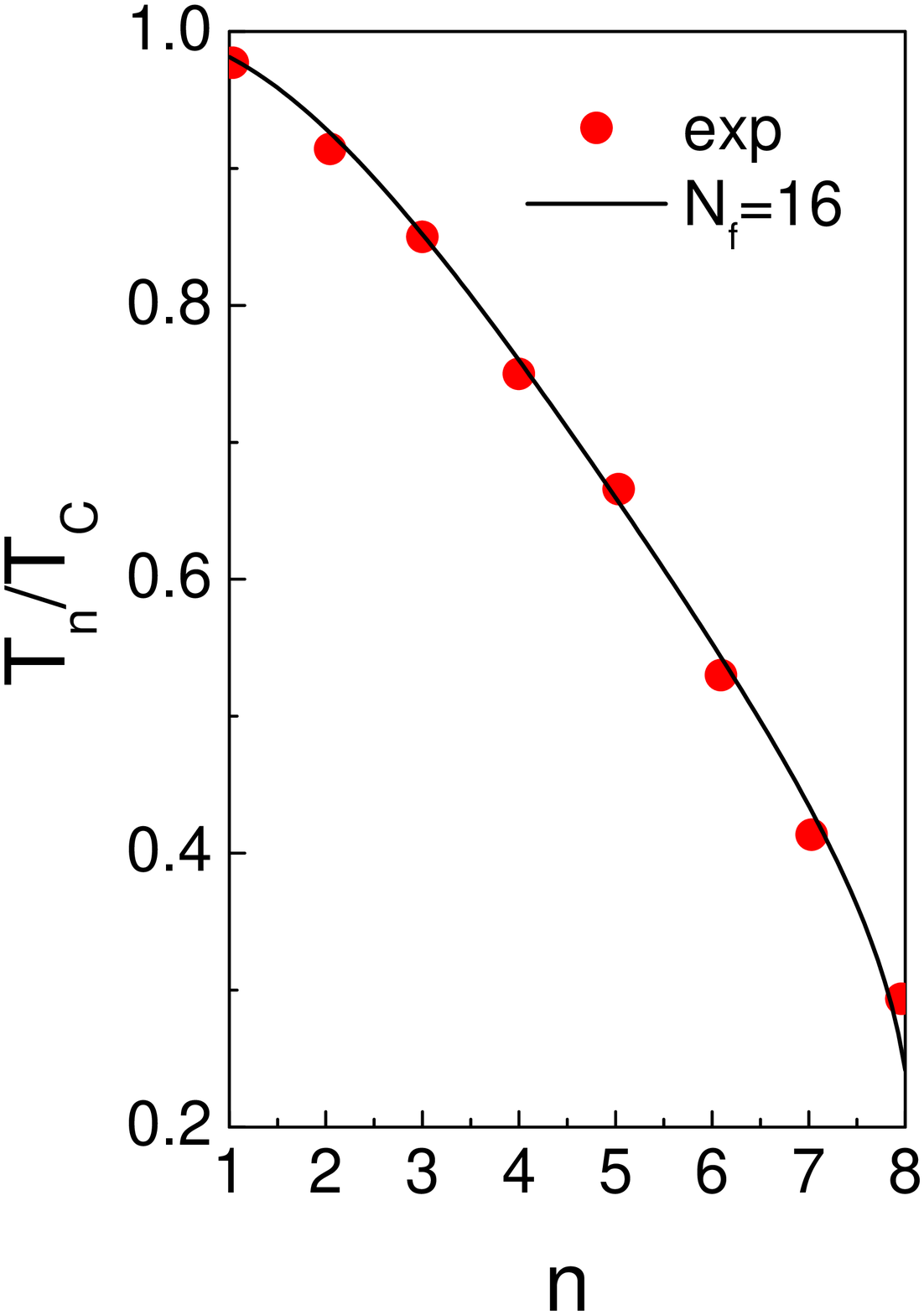}}\vspace{0.25cm}
\caption{The predicted dependence (solid line) of the reduced
temperature $T_n/T_C$ on the number of oscillations minima $n$ for
$N_f=16$ (according to Eq.(4)) along with the extracted (from
Fig.\ref{fig:fig2}) experimental points.} \label{fig:fig5}
\end{figure}

The best fits of the high-temperature data, using Eqs.(2) and (3),
produced $N_f=L/l=16$ for the number of particles contributing to
the observed oscillating behavior of nanofluid magnetization
(which reasonably correlates with an average cluster size of
$L=160nm$, revealed by the TEM images of the nanofluid, see
Fig.\ref{fig:fig1}).

Notice also that, according to Eq.(3), the number of oscillation
minima $n$ of magnetization (observed at $T=T_n$) is given by the
solution of the implicit equation, $k(T_n)L=2\pi n$ where
$n=1,2,3,..$. Using the Curie-Weiss expression for bulk
magnetization $M_{\infty}(T)$, the above equation results in the
following explicit dependence of $T_n$ on $n$ and $N_f$:
\begin{equation}
T_n=\frac{T_C}{\sqrt{1+\{\tanh ^{-1}(2n/N_f)\}^2}}
\end{equation}

Fig.\ref{fig:fig4} demonstrates very good agreement between the
predicted $n$ dependence of $T_n/T_C$ (given by Eq.(4) with
$N_f=16$) and the extracted from Fig.2 experimental points.

In summary, we reported the magnetic properties of the recently
synthesized nanofluid room-temperature ferromagnetic graphite
(with the single particle size of $l\simeq 10nm$). In addition to
a low-temperature magnetic anomaly (attributed to the
manifestation of quantum size effects below $50K$), we also
observed strong temperature oscillations of spontaneous
magnetization  (attributed to manifestation of the hard-sphere
type pair correlations between ferromagnetic particles in the
nanofluid above $50K$).

\ack This work has been financially supported by the Brazilian
agencies CNPq, CAPES and FAPESP.

\end{document}